\documentclass[12pt, a4paper]{elsarticle}

\usepackage{xargs}                      
\usepackage[pdftex,dvipsnames]{xcolor}  
\usepackage[colorinlistoftodos,prependcaption]{todonotes}
\usepackage[utf8]{inputenc}
\usepackage{amsmath}
\usepackage{amssymb}  
\usepackage{amsthm}
\usepackage{textcomp}
\usepackage{enumerate}
\usepackage{array}
\usepackage{bbm}
\usepackage{verbatim}
\usepackage{hyperref}
\usepackage{graphicx}
\usepackage{subfigure}
\usepackage{mathtools}
\usepackage{cleveref}

\newcommand{\trial}[0]{\ensuremath{\mathrm{\operatorname{trial}}}}

\newcommand{\V}    {\operatorname{Var}}
\newcommand{\err}  {\operatorname{error}}

\renewcommand{\P}{\mathbb{P}}
\newcommand{\E}{\mathbb{E}}
\newcommand{\MC}{\mathrm{MC}}
\newcommand{\MLMC}{\mathrm{MLMC}}

\newcommand{\cost} {\operatorname{cost}}
\newcommand{\red}[1]{\textcolor{black}{#1}}
\newtheorem{theorem}{Theorem}

\DeclareMathOperator*{\argmax}{arg\,max}

\theoremstyle{definition}

\theoremstyle{remark}
\newtheorem{remark}[theorem]{Remark}

\newcommand\numberthis{\addtocounter{equation}{1}\tag{\theequation}}

\newcommand\revision[1]{{\color{black} #1}}

\title{Multilevel Monte Carlo for Reliability Theory}
\author[ox]{Louis~J.~M.~Aslett\corref{cor}}
\ead{aslett@stats.ox.ac.uk}
\author[ox]{Tigran~Nagapetyan}
\ead{nagapetyan@stats.ox.ac.uk}
\author[ox]{Sebastian~J.~Vollmer}
\ead{vollmer@stats.ox.ac.uk}

\cortext[cor]{Corresponding author}

\address[ox]{Department
of Statistics, University of Oxford, 24--29 St Giles',
Oxford, OX1 3LB, United Kingdom}

\begin{document}

\begin{abstract}
As the size of engineered systems grows, problems in reliability theory can become computationally challenging, often due to the combinatorial growth in the number of cut sets.  In this paper we demonstrate how Multilevel Monte Carlo (MLMC) --- a simulation approach which is typically used for stochastic differential equation models --- can be applied in reliability problems by carefully controlling the bias-variance tradeoff in approximating large system behaviour.  In this first exposition of MLMC methods in reliability problems we address the canonical problem of estimating the expectation of a functional of system lifetime \red{for non-repairable and repairable components, demonstrating} the computational advantages compared to classical Monte Carlo methods.  The difference in computational complexity can be orders of magnitude for very large or complicated system structures\red{, or where the desired precision is lower}.
\end{abstract}


\maketitle

\section{Introduction}

It can prove to be computationally intractable to perform classical reliability analysis of very large engineered systems when the number of cut (path) sets
grows combinatorially.  It is well understood that working instead with subsets of the cut (path) sets or bounding structural designs can provide probability bounds in many reliability
problems \citep{Barlow1965}, but such bounds can be crude or may not be well characterised at all.

Evaluation of the reliability of engineered systems is a crucial part
of system design and often scenario planning may involve
repeated evaluation of the reliability for changing system 
configurations or component types meaning rapid simulation is highly desirable.
For simplicity of exposition we herein consider the canonical problem of
estimating the expectation of a functional of system lifetime \red{both with and without a component repair process, showing} the approach 
developed is easily generalised
to other reliability problems which depend on cut (path) sets for the analysis.

\revision{In the case of static reliability analysis, there are many methods aside from Monte Carlo simulation using the cut (path) sets, including 
Sum of Disjoint Products (SDP) methods \citep{Rai:1995aa, zuo2007efficient, yeh2015improved} and methods based on Binary Decision Diagrams 
(BDD) \citep{xing2015binary} or multistate BDD extensions \citep{zang2003bdd}.  On the other hand, these approaches are less prevalent in dynamic 
reliability problems where there are component dependencies, for example through system shocks, repair or maintenance programmes, and cascading failures 
among others.  There have been recent developments in dynamic fault trees \citep{Merle:2010aa,xing2011exact,merle2014quantitative} which apply where 
event sequence ordering influences the reliability, including repairable systems \citep{manno2014conception}.  When there are arbitrary dependencies, 
the most generally applicable approach is direct Monte Carlo simulation (e.g. \citep{chiacchio2016stochastic}), so that acceleration of Monte Carlo 
techniques is important to address a broad range of the most complex reliability scenarios.  Monte Carlo acceleration through importance sampling \citep{jun1992system}, or the use of control variates \citep{yevkin2010improved} have been suggested 
in the context of reliability estimation, but they are either restricted to the static case and require regular updates and sorting of all the cut sets (as for \citep{jun1992system}), or could be combined with the MLMC paradigm (as for \citep{yevkin2010improved}).}

\revision{Indeed, also note that interest may not be in the reliability at a particular fixed mission time, but instead in: some expectation of a functional of system lifetime; or in ascertaining a quantile of system lifetime (i.e.~the time to which one is 99.9\% certain the system will survive); or in estimation of the entire system lifetime distribution.  In these situations Monte Carlo methods are typically the only tractable approach.}


Multilevel Monte Carlo (MLMC) methods --- pioneered by Heinrich 
\citep{heinrich98} and Giles \citep{giles08} --- are now standard for 
estimation of expectations of functionals of processes defined by
stochastic differential equations (SDEs).  \revision{However, the MLMC approach is in fact 
a general paradigm for accelerating any Monte Carlo based method (whether 
standard, importance sampling, etc), if one can link the accuracy of the 
estimator with the complexity of generating a sample, while at the same time controlling the variance of the difference for approximations with different accuracy.}
\revision{The main contribution in this paper is development of a Multilevel Monte Carlo (MLMC) approach to reliability problems.}
\revision{In this way, we show how any reliability problem using Monte Carlo simulation over cut (path) sets can be substantially accelerated}, 
extending the size of systems \red{and complexity of dependencies} which are within reach for reliability evaluation.

In Section \ref{sec:SimSysLife}, we recap the traditional cut set method of simulating system lifetimes which does not scale well to large systems 
even when the cut sets are known.  This motivates the approach taken in this work.
In Section \ref{sec:mc} we recap standard Monte Carlo theory and set out the error and computational cost metrics which will enable comparison 
with our MLMC based approach.
The fundamental MLMC methodology and our adaptation to the reliability setting then follow in Section \ref{sec:mlmc}, before numerical results 
demonstrating the kind of substantial computational improvements which can be achieved are covered in \Cref{sec:examples,sec:examplesrepair}.

%

\section{Simulating system lifetimes}\label{sec:SimSysLife}
Consider a coherent system with $n$ components.  Let $x_1(t), \dots, x_n(t)$ denote the operational status ($1=$ working, $0=$ failed) of the components at time $t$ and consider the random variable for the lifetime of component $c$ to be $T_c \sim F_c(t)$, where $F_c(\cdot)$ are positively supported lifetime distributions which are not necessarily independent or identical.
We will depict a system as an undirected network comprising a set of nodes (vertices) $S$, and a set of edges $E$, where nodes are considered unreliable and edges are perfectly reliable (note that any setting with imperfect edge reliability can be easily transformed to a corresponding representation where they are perfectly reliable \citep{acw15}).
\begin{figure}
  \centering
    \includegraphics[width=0.45\textwidth]{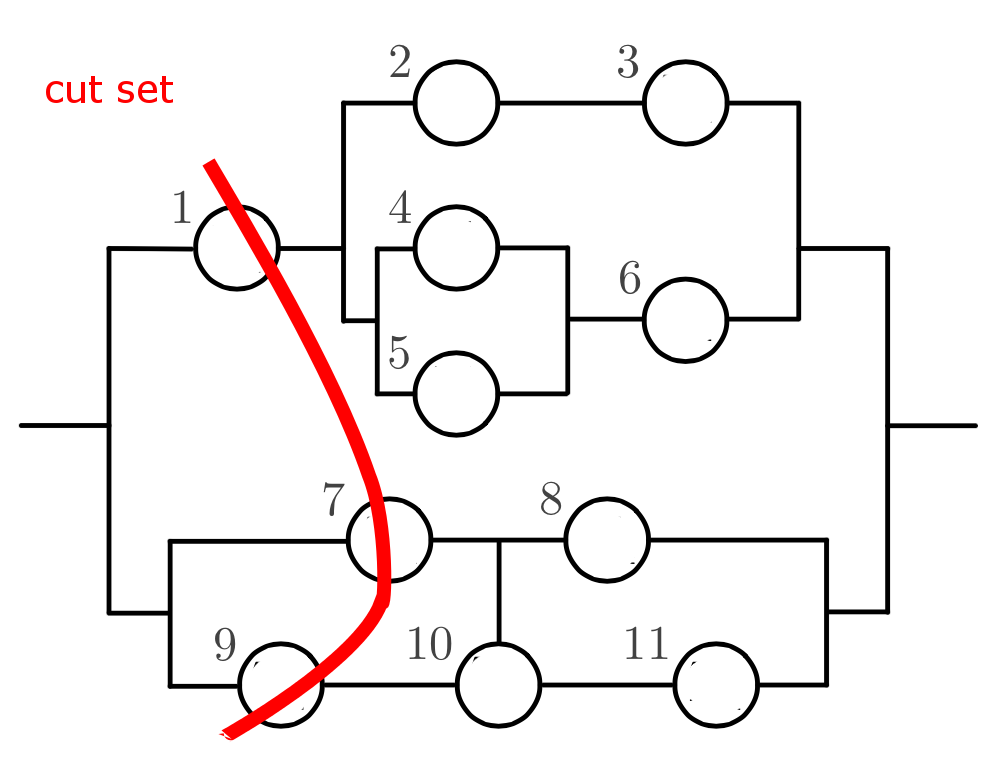}
      \caption{A sample network with a minimal cut set.}
      \label{Fig1}
\end{figure}
The system is considered to be functional as long as there is a path from left to right which passes only though functioning nodes, see Figure \ref{Fig1}.  This is usually represented mathematically by the structure function, $\phi : \{0,1\}^n \to \{0,1\}$, which maps component status to system status.

Herein, our focus is on an equivalent means of evaluation based on cut sets.  A set of components, $\mathrm{C}$, is said to be a cut set of the system if the system is failed whenever all the components in $\mathrm{C}$ are failed.  A cut set is said to be a minimal cut set if no subset of it is also a cut set.  Then, the set of all minimal cut sets, $\mathcal{C}$, characterises the operational state of a system completely and is equivalent to knowledge of the structure function \citep{Butterworth1972}.  In addition to the cut sets characterising the operational state of the system given the binary operational state of the components, they also immediately provide the system failure time if the individual component failure times are known \citep{Barlow1981}:
\[ T_S = f_S(T_1, \dots, T_n) := \min_{\mathrm{C} \in \mathcal{C}} \left\{ \max_{c \in \mathrm{C}} \{ T_c \} \right\}. \]
Thus, the failure time for the system depends on the system structure and the failure time distributions for each node.

The traditional approach to estimating the expectation of a functional of the lifetime of a system given the lifetime distributions of the components is to perform a simple Monte Carlo simulation.  That is,
\[ \E[g(T_S)] = \frac{1}{N} \sum_{i=1}^N g(f_S(t_1^{(i)}, \dots, t_n^{(i)})) \mbox{ where } t_j^{(i)} \sim F_j(\cdot). \]

The overall runtime for this approach depends on three quantities:
\begin{enumerate}
\item Variance of the estimator. Due to the random nature of component failure times, the estimator is a random variable: higher variance estimators will require more iterations to achieve an accurate estimate;
\item Target accuracy of the estimate. Naturally, the higher the desired accuracy, the longer the algorithm will take due to more iterations being required;
\item Number of cut sets.  As the system size grows, the number of cut sets has a combinatorial growth, making the approach impractical for very large systems.
\end{enumerate}

Less brute force approaches are possible with the restrictive assumption of iid components by making use of the system signature \citep{Kochar1999,Samaniego1985}.  More recent work on the survival signature \citep{Coolen2012} has generalised the signature to multiple types of component, with the weaker assumption of exchangeability between components.  However, if a large number of the components are of different types or there are highly dependent failures, then the survival signature will also grow exponentially in complexity.  It can also accommodate a repair process \citep{Coolen:2014aa} through expression as a new component type, though this increases the complexity if too many repairs occur.  Hence, in this work, we first address the most general possible setting in which any form of component lifetime and dependence structure is allowed, requiring only knowledge of component lifetimes and the cut sets of the system.  However, note that it should be possible to specialise this approach to work with the survival signature which we hope to address in future research.

\section{Monte Carlo algorithms}\label{sec:mc}
To simplify presentation, hereinafter we only consider estimating expected failure time directly, rather than some functional of the failure time, though this is mostly without loss of generality (see \Cref{sec:mlmc} for details). 
Therefore, assume that for a given system $S$, we want to estimate the expected failure time.
\[\E T_S =  \E f_S(t_1^{(i)}, \dots, t_n^{(i)}).\]

There are many approaches to simulation which may differ in terms of convergence to the true value as well as computational characteristics.  In order to compare them, we present some useful cost and error expressions in the following subsection.

\subsection{Performance measures: error and cost definitions}
We start by defining the two main quantities, which will be used throughout this paper to compare methodologies.
Given an estimator $\hat{T}_S$ of the quantity $\E T_S$, the Mean Squared Error (MSE) of any Monte Carlo based method is:
$$\err = \E\left[\left((\hat{T}_S) - \E T_S\right)^2\right].$$
The classical decomposition of the MSE yields:
\begin{gather*}
\E\left[\left(\hat{T}_S - \E T_S\right)^2\right] = \E\left[\left(\hat{T}_S + \E \hat{T}_S - \E \hat{T}_S- \E T_S\right)^2\right]\\
 = \E\left[\left(\hat{T}_S - \E \hat{T}_S\right)^2\right] + \left(\E\hat{T}_S - \E T_S\right)^2\numberthis \label{eq:mse}
\end{gather*}
where $\left(\E\hat{T}_S - \E T_S\right)^2$ is the squared bias error, while $\E\left[\left(\hat{T}_S - \E \hat{T}_S\right)^2\right]$ is the error due to Monte Carlo variance. The first is a systematic error arising from the fact that we might not sample our random variable exactly, but rather use a suitable approximation, while the second error comes from the randomised nature of any Monte Carlo algorithm.  For example, in traditional Monte Carlo applications, one samples exactly so that the first error is zero and only the Monte Carlo variance needs to be treated carefully.

The cost of any Monte Carlo based algorithm is typically taken to be the expected 
runtime in order to achieve a prescribed accuracy.  A more convenient approach 
for theoretical comparison between different methods is to define
$$\cost = \E(\text{\#random number generations and operations}).$$

We now recap traditional Monte Carlo and then introduce Multilevel Monte Carlo, 
in each case highlighting their corresponding results for these two measures of performance.

\subsection{Traditional (or single-level) Monte Carlo algorithm}
The traditional Monte Carlo estimator is based on $N$ replications of simulating 
the system lifetime, via the minimal cut sets, by simulating the component 
lifetimes. That is, given system simulations $\tau_i = f_S(t_1^{(i)}, \dots, t_n^{(i)})$
the traditional Monte Carlo estimator has the form
\begin{equation}
\label{eq:smc_est}
\hat{T}_S = \cfrac1N\sum\limits_{i=1}^N f_S(t_1^{(i)}, \dots, t_n^{(i)}).
\end{equation}
For reasons that will become clear in the sequel, it is useful to refer to this 
as the single-level Monte Carlo algorithm because it emphasises the relationship 
to Multilevel Monte Carlo.

This single-level Monte Carlo estimate has variance proportional to $N^{-1}$,
\[
\V(\hat{T}_S) = \V\left(\frac1N\sum\limits_{i=1}^N \tau_i\right)=\frac{1}{N^2}\V\left(\sum\limits_{i=1}^N \tau_i\right)=\frac1N\V\left(\tau_i\right).
\]
The estimator \eqref{eq:smc_est} is clearly unbiased, because there is no 
approximation involved in estimating the failure time, so the error measure 
introduced earlier only has this second variance term, \[ \err_\MC = N^{-1} \V\left(\tau_i\right). \]  Indeed, more generally 
the well known central limit result for standard Monte Carlo means that:

\[ \mathbb{P}\left( |\hat{T}_S - \E T_S| > z\frac{\sqrt{\V(\tau_i)}}{\sqrt{n}} \right) \approx \mathbb{P}(|Z| > z) \]
for \(Z \sim \mathrm{N}(0,1)\).

Thus, for a desired level of accuracy \(\varepsilon > 0\) with confidence level
\(1-\alpha\), we require
\[ n = z_{\alpha/2}^2 \V(\tau_i) \varepsilon^{-2} \] Monte Carlo
simulations, where the quantile \(z_{\alpha/2}\) is chosen to ensure that
\(\mathbb{P}(Z > z_{\alpha/2}) = \alpha/2\).

Naturally \(z_{\alpha/2}\) is a constant for any fixed level of confidence, so 
the variable compute costs in simulation are
\begin{equation}
\label{eq:cost:mc}
\cost_\MC = \V(\tau_i)\cdot \varepsilon^{-2} \cdot \#\mathcal{C},
\end{equation}
where $\#\mathcal{C}$ denotes the number of minimal cut sets for the system.

\section{Multilevel Monte Carlo}\label{sec:mlmc}
To simplify presentation we again only consider estimating expected failure time 
directly, rather than some functional of the failure time.  Note that there is 
no loss of generality, so long as the functional of interest is Lipschitz 
continuous (or bounded for discrete measures). The most common functional of 
interest that this would exclude is computing expectations of quantiles.  
However, this problem can be treated with a smoothing approach, as discussed for 
the MLMC setting in \cite{gnr15}.  In all other cases, the presentation below 
carries over in the natural fashion.

\subsection{General MLMC}
We first introduce MLMC in generality before specialising this to the reliability problem considered herein.
Consider a sequence of estimators $T_0, T_1, \ldots,$ which approximates $T_L$ with 
increasing accuracy, but also increasing cost.  By linearity of expectation, we have
\[
\E(T_L) = \E(T_0) + \sum_{\ell=1}^L \E[T_\ell-T_{\ell-1}],
\]
and therefore we can use the following unbiased estimator
for $\E[T_L]$,
\[
\frac{1}{N_0} \sum_{n=1}^{N_0} T_0^{(0,n)} \ + \ 
 \sum_{\ell=1}^L \left\{
\frac{1}{N_\ell} \sum_{n=1}^{N_\ell} \left(T_\ell^{(\ell,n)} - T_{\ell-1}^{(\ell,n)}\right)
\right\}
\]
The inclusion of the level $\ell$ in the superscript $(\ell,n)$ 
indicates that the samples used at each level of correction are 
independent, but crucially note that the differences themselves
use common samples.  Note the terminology `correction' arises from the fact
that each $T_\ell$ is generally \emph{not} an unbiased estimate any more.

Let $V_0$ and $\cost_0$ be the variance and the expected cost of one sample 
of $T_0$, and let $V_\ell,\ \cost_\ell$ be the variance and expected cost of one 
sample of $T_\ell-T_{\ell-1}$.  Then the overall expected cost and variance
of the multilevel estimator are $\sum_{\ell=0}^L N_\ell\cdot \cost_\ell\ $ and $\sum_{\ell=0}^L N_\ell^{-1} \cdot V_\ell$, respectively.

More generally, this means that provided the product $V_\ell \cdot \cost_\ell$ 
decreases with $\ell$ (i.e.~the cost increases with
level slower than the variance decreases), then one can achieve significant computational savings, which can be formalised in Theorem \ref{thm:compl} from \cite{giles08}.

\begin{theorem}
\label{thm:compl}
Let $T_S$ denote a random variable, and let $T_\ell$ denote the 
corresponding level $\ell$ numerical approximation.

If there exist independent estimators $Y_\ell$ based on $N_\ell$ 
Monte Carlo samples, and positive constants 
$\alpha, \beta, \gamma, c_1, c_2, c_3$ such that 
$\alpha\geq{\textstyle \frac{1}{2}}\,\min(\beta,\gamma)$ and
\begin{enumerate}
\item 
$
\left| \E(T_\ell - T_S) \right|\ \leq\ c_1\, 2^{-\alpha\, \ell}
$
\item
$
\E(Y_\ell)\ = \left\{ \begin{array}{ll}
\E(T_0),                     &~~ \ell=0 \\[0.1in]
\E(T_\ell - T_{\ell-1}), &~~ \ell>0
\end{array}\right.
$
\item
$
\V(Y_\ell)\ \leq\ c_2\, N_\ell^{-1}\, 2^{-\beta\, \ell}
$
\item
$
\cost_\ell\ \leq\ c_3\,2^{\gamma\, \ell},
$
where $cost_\ell$ is the expected computational complexity of  $Y_\ell$
\end{enumerate}
then there exists a positive constant $c_4$ such that for any 
$\varepsilon < e^{-1}$
there are values $L$ and $N_\ell$ for which the multilevel estimator
\[
Y = \sum_{\ell=0}^L Y_\ell,
\]
has a mean-square-error with bound
\[
MSE \equiv \E\left[ \left(Y - \E[T_S]\right)^2\right] < \varepsilon^2
\]
with a computational complexity $C$ with bound
\[
\cost_{MLMC} \leq \left\{\begin{array}{ll}
c_4\, \varepsilon^{-2}              ,    & ~~ \beta>\gamma, \\[0.1in]
c_4\, \varepsilon^{-2} (\log \varepsilon)^2,    & ~~ \beta=\gamma, \\[0.1in]
c_4\, \varepsilon^{-2-(\gamma-\beta)/\alpha}, & ~~ \beta<\gamma.
\end{array}\right.
\]
\end{theorem}

\begin{remark}
We will informally illustrate the idea behind MLMC on a simple example with just two levels.
Consider just two approximations $T_k$ and $T_L$, where $k<L$, with sample costs $\cost_k<\cost_L$. It is clear, that the cost of one sample for $T_L - T_k$ is roughly $\cost_L$. Now assume, that
$$V_1=\V T_k\approx \V T_L, \text{ and } V_2=\V (T_L - T_k),$$
where $V_2<V_1$.
Then we have 
\begin{align*}
\E T_L &= \E T_k + \E (T_L - T_k)\\
\Rightarrow \hat{T}&=\frac1N \sum_{n=1}^{N} T_L^{(2,n)}\\
&\approx \bar{T}=\frac1N_1 \sum_{i=1}^{N_1} T_k^{(1,i)} + \frac1N_2 \sum_{j=1}^{N_2} (T_L^{(1,i)}-T_k^{(1,i)}).
\end{align*}
We see that the overall cost of Monte Carlo estimators, according to \eqref{eq:cost:mc}, can be expressed as
\begin{gather*}
\cost(\hat{T}) = \varepsilon^{-2}\cdot V_1\cdot \cost_L\\
\cost(\bar{T}) = \varepsilon^{-2}\cdot \left(\cost_k\cdot V_1 + \cost_L\cdot V_2\right),
\end{gather*}
which gives us a condition
$$\cost(\hat{T}) > \cost(\bar{T})\Rightarrow 1> \frac{\cost_k}{\cost_L} + \frac{V_2}{V_1}.$$
In other words, provided there exists a good coupling between estimators $T_k$ and $T_L$, we have reduced computational cost even for two levels. The two-level Monte Carlo method in the context of Monte Carlo path simulation has been suggested and analysed in\cite{kebaier05}.
\end{remark}

\begin{remark}
  \red{Multilevel provides the greatest benefit when $\beta \ge \gamma$, because this is the case for which we get the best asymptotic performance. $\gamma$ represents the parameter for the exponential increase of the cost of producing a sample, while $\beta$ corresponds to the parameter for the exponential decay in the variance of the sample at a given level.  There are 3 cases:}
  \red{
\begin{enumerate}
\item[$\beta<\gamma$.] The number of samples required by the MLMC estimator decays at a slower rate than the increase of the sampling cost at each level. In this case the overall cost is proportional to the cost of the last level.
\item[$\beta=\gamma$.] This is most common in practice \cite{giles15}. The decay of the variance is balanced with the increase of the cost, therefore the contribution to the overall cost is the same from all the levels.
\item[$\beta>\gamma$.] In this most desirable case, the overall cost is dominated by level $l=0$, since consecutive levels will have a decaying contribution to the cost.
\end{enumerate}}
  \red{When not available analytically, estimation of $\alpha, \beta$ and $\gamma$ is usually done by empirically regressing using diagnostic quantities in the manner demonstrated in \Cref{sec:examples,sec:examplesrepair} for our examples.}
\end{remark}

\begin{remark}
Multilevel Monte Carlo became popular after the seminal work of Mike Giles \cite{giles08} for estimating expectations of functionals $\E(f(X_t))$, where $X_t$ is the solution of a stochastic differential equation. In the general Multilevel Monte Carlo path simulation setting, $T_{\ell}$ from Theorem \ref{thm:compl} is the functional value, evaluated via an approximation arising from a discretisation method, e.g. the Euler-Maruyama method \cite{kloeden1992numerical}.
\end{remark}
\subsection{MLMC for system reliability}\label{sec:RelMLMC}
Theorem \ref{thm:compl} suggests that one may want to try getting a coarser Monte Carlo estimate of the system lifetime, perhaps by considering only a subset of the collection of minimal cut sets.
\[ \mathcal{C}' \subset \mathcal{C} \implies \min_{C \in \mathcal{C}'} \big\{ \max_{i \in C} \{ t_i \} \big\} = T'_S \ge T_S = \min_{C \in \mathcal{C}} \big\{ \max_{i \in C} \{ t_i \} \big\}. \]

On its own $T'_S$ is a biased estimator, so although a traditional single-level Monte Carlo estimator based on it may have lower computational cost, it will have increased MSE because the first error term in \eqref{eq:mse} can no longer be ignored.
However, by using this coarse estimate inside an MLMC approach, we aim to improve the overall performance. To this end we introduce the sequence of estimators \(T_0, \dots, T_L\) based on a nested sequence of
minimal cut sets,
\(\mathcal{C}_0 \subset \dots \subset \mathcal{C}_L = \mathcal{C}\).  Note that here $T_L \equiv T_S$, which is not typically true in a general MLMC setting.

The crucial ingredient is the finite telescopic sum
\[ \E(T_S) \equiv \E(T_0) + \sum_{\ell=1}^L \E(T_\ell - T_{\ell-1})= \sum_{\ell=0}^L\E(Y_\ell) \]

As described above, we independently estimate each term, and within each term, \(T_\ell\)
and \(T_{\ell-1}\) use the same random component simulations:
\[ \E(Y_\ell)\approx N_\ell^{-1} \sum_{j=1}^{N_\ell} \Big(\tau_\ell^{(j)} - \tau _{\ell-1}^{(j)}\Big) \]
with each level having cost being bounded from above by $c\cdot \V(Y_\ell)\cdot \#\mathcal{C}_\ell$. Here $c$ is a constant independent of $\ell$ and the desired target accuracy. We choose $\#\mathcal{C}_\ell$ --- the number of minimal cut sets at level $\ell$ --- to be a proxy for the upper bound on the cost of each level, because for a fixed system the number of elements in each minimal cut set is independent of the target accuracy.  In other words,  as we double the number of minimal cut sets in each level, their number is a straightforward way to construct a meaningful and easy upper bound for the cost of one sample.\label{txt:upperbd}

Thus, the overall MLMC variance is
\[ \mbox{Var}\left( \sum_{\ell=0}^L Y_\ell \right) = \sum_{\ell=0}^L N_\ell^{-1} \V(Y_\ell) \]
at a cost of $\sum_{\ell=0}^{L} N_\ell\cdot \#\mathcal{C}_\ell$.  Therefore, given a fixed target accuracy (variance), if we choose a sample size
\(N_\ell \approx \sqrt{\V(Y_\ell) / \#\mathcal{C}_l}\) on each level, this will minimise
the computational cost.
That is, for a desired accuracy $\varepsilon > 0$, the overall
cost is:

\begin{equation}  
  \mbox{Cost}_\MLMC\ = \sum_{\ell=0}^{L} N_\ell \cdot\# \mathcal{C}_\ell = \varepsilon^{-2} \left( \sum_{\ell=0}^{L} \sqrt{\V(Y_\ell) \cdot \#\mathcal{C}_\ell} \right)^2
  \label{eq:cost:mlmc}
\end{equation}

Recall that,
\[ \cost_\MC = \V(\tau_i)\cdot \varepsilon^{-2} \cdot \#\mathcal{C}, \]
This means that:
\begin{enumerate}
\item If we have a good coupling between the approximations, or equivalently $\V(Y_\ell)$ decays rapidly, then we can achieve considerable savings compared to the single-level Monte Carlo algorithm.
\item Additional savings are possible if we do not calculate all the levels $Y_\ell,$ but rather stop the algorithm early.  This introduces a (small) bias, but substantially decreases the overall computational cost.  As long as the bias is quantified --- and when combined with the estimator variance gives a MSE \eqref{eq:mse} below our target accuracy --- then we can still solve the original problem at much lower cost.

    Our proposed application to reliability involves a nested sequence of minimal cut sets providing improving accuracy:
\[ \mathcal{C}_0 \subset \dots \subset \mathcal{C}_L = \mathcal{C}, \]
so the possible gain from stopping early depends on the way the minimal cut sets are chosen at each level.
\end{enumerate}

\subsubsection{Level selection algorithm}
\label{sec:levelselection}

The first point to note is that existing Multilevel Monte Carlo literature has 
shown that anything less than geometric decay in the cost of computation at each 
level leads to suboptimal gains, see \cite{giles15}.  Therefore, we pre-specify that level $\ell$ 
contain $\lceil \#\mathcal{C}/2^{L-\ell} \rceil$ minimal cut sets.  The levels will be grown 
from $\ell=0$ up, adding in those minimal cut sets which are in some sense 
optimal for the next level.  Thus, we specify level selection in an inductive 
fashion.  \red{Note, we initially ignore the possibility of repair for simplicity of presentation and address
the changes involved to accommodate repair in \Cref{sec:repair_lvlsel}, when we demonstrate MLMC for repairable components.}

\subsubsection*{Selection of level $\ell=0$}

Level 0 will be simulated most frequently since it is the lowest cost.  
Therefore, optimal choice of this level is straightforward: it should contain 
the minimal cut sets which provide the best approximation to $T_S$.  That is, we 
wish to assign to level 0 the minimal cut sets which have smallest expected 
failure time, since these will most frequently be the causes of system failure.  
To achieve this, we propose an initial highly crude estimate by performing a 
pilot standard Monte Carlo simulation of $N'$ lifetimes of each component in the 
system, using these to generate $N'$ realisations of the failure time associated 
with each minimal cut set,
\begin{equation}
  \eta_{i} = \frac{1}{N'} \sum_{j=1}^{N'} \max_{c \in C_i} \left\{ t_c^{(j)} \right\}, \quad \forall\ C_i \in \mathcal{C} .
  \label{eq:lvl0}
\end{equation}
The cut sets corresponding to the smallest $\lceil \#\mathcal{C}/2^L \rceil$
of these $\eta_i$ are then chosen to form $\mathcal{C}_0$.

\subsubsection*{Selection of level $\ell>0$}

Given that we have chosen levels $0, \dots, \ell-1$ already (that is $\mathcal{C}_0 \subset \dots \subset \mathcal{C}_{\ell-1}$ are now fixed), we need to select which cut sets to add from $\mathcal{C}_{\trial} = \mathcal{C} \backslash \mathcal{C}_{\ell-1}$.  To maximise the performance of MLMC we would like to select the cut sets such that
$$\E [T_{\ell-1} - T_{\ell}]\to\max.$$
In other words, choose
$$\mathcal{C_\ell} = \argmax_{\mathclap{\substack{\mathcal{C_\ell} \subseteq \mathcal{C} \\ \mathrm{s.t.}\  \mathcal{C}_{\ell-1} \subset \mathcal{C_\ell}, \#\mathcal{C}_\ell = \lceil \#\mathcal{C}/2^{L-\ell} \rceil}}} \ \E [T_{\ell-1} - T_{\ell}],$$
so that the contribution from each level is as large as possible in the smallest 
levels, leading to a rapid decay in the size of the contribution in each level and hence the possibility
of terminating the algorithm early.  In particular, note that if 
$\sum_{\ell=k}^L \E(T_\ell - T_{\ell-1}) \ll \varepsilon$, then levels 
$k, \dots, L$ need not be simulated at all, so that ensuring all large 
differences occur in early levels is highly desirable.

Notice that:
$$\E [T_{\ell-1} - T_{\ell}]\le \E\left[T_{\ell-1} - \min\left\{T_{\ell-1},\max\limits_{c \in C}\left\{t_c\right\} \right\}\right],$$
for any $C \in \mathcal{C}_{\trial}$.  So our choice for sorting cut sets is motivated by the minimisation of the upper bound for the increments at each level, which we can implement for any level $\ell$ in a simple way:
\begin{itemize}
\item use the $N'$ samples and calculated failure times for all cut sets used in selecting $\ell=0$,
 \item calculate the following estimates:
 $$\delta_k = \frac{1}{N'}\sum_{j=1}^{N'} \left[T_{\ell-1} - \min\left\{T_{\ell-1},\max\limits_{c \in C_k}\left\{t_c^{(j)}\right\} \right\}\right],$$
  for each $C_k \in \mathcal{C}_{\trial}$.
\item Sort $\delta_k$ in a descending order and add cut sets corresponding to the largest values for $\delta_k$ to $\mathcal{C}_\ell$ until $\#\mathcal{C}_\ell = \lceil \#\mathcal{C}/2^{L-\ell} \rceil$.
\end{itemize}
This choice of number of minimal cut sets on each level guarantees an exponential increase in the cost, corresponding to $\gamma=1$ in Theorem \ref{thm:compl}.

\subsection{Full Multilevel Monte Carlo algorithm for reliability}
One of the key features of the Multilevel Monte Carlo algorithm is its ability to naturally provide stopping criteria for an optimal selection of the number of levels $\hat{L}$ to actually simulate, which we illustrate now along with the full description of the algorithm. For more advanced approaches to implementation of Multilevel Monte Carlo we refer to \cite{giles_MLMC_code, Aslett:2016aa, giles15}.

According to Theorem \ref{thm:compl} and the first assumption in it, we have asymptotically as $\ell\to\infty$
$$\E(Y_\ell)\approx \E(T_S-\hat{T}_{\ell}),$$
so that a natural stopping criteria is to choose $\hat{L}$ minimal such that $|Y_{\hat{L}}| \leq \varepsilon/2$.
\begin{enumerate}
 \item[Input:] Required accuracy $\varepsilon$ and level specification as per \S\ref{sec:RelMLMC}.
 \item Set the initial number of levels to $\hat{L}:=2$. In order to define the optimal number of samples at each level we need to estimate the variances on levels $\ell=0,1,2$
 \item Compute $N_{\ell}:=100$ samples on levels $\ell=0,1,2$
\item Estimate $\V(Y_\ell)$ and update $N_{\ell}$ for each level $\ell=0,\dots,\hat{L}$:
   \begin{gather*}
   N_\ell := \max\left\{N_\ell, \hat{N}_\ell\right\},\text{ where}\\
   \hat{N}_\ell=\left\lceil 4\cdot\varepsilon^{-2} \cdot \sqrt{\V(Y_\ell)  2^{-\ell}} \cdot \sum_{k=0}^{\hat{L}} \sqrt{\V(Y_k)\cdot 2^{k}}\right\rceil
   \end{gather*}
We take the maximum as it may happen that the numerical variance was initially overestimated and so more simulations were performed than necessary. If $N_\ell$ has increased less than 1\% on all levels, then skip to step 5.
\item Compute the additional number of samples on each level $\ell=0, \dots, \hat{L}$ and return to step 3.
\item Upon reaching this step we have converged in terms of MSE due to the variance.  Next we test whether the bias error term is sufficiently small to terminate, or whether more levels and simulations are required. If $|Y_\ell|\geq \varepsilon$:\\
Then: set $\hat{L}:=\hat{L}+1$ , $\V(\hat{Y}_{\hat{L}}) =\V(Y_{\hat{L}-1})/ 2$ and return to step 3;\\
Else: Terminate algorithm returning $\sum_{\ell=0}^{\hat{L}} Y_\ell$ as the estimate.
\end{enumerate}
There are two extreme cases to bear in mind. In the first case, we have only a few minimal cut sets (or even only one), which influence the failure time. This case is treated with the initial choice of the cut sets and selection as prescribed in \S\ref{sec:RelMLMC} should ensure the minimal number of levels is simulated.  The second case is when all the cut sets have similar `weight' in determining the failure time, such as with a fully connected system with independent and identically distributed failure times for all components.  This case is treated with the doubling of the number of cut sets with respect to the previous level, which again assures the mean and variance decay between the levels.
\begin{remark}
The Multilevel paradigm, with some slight modifications, also allows construction of efficient numerical algorithms (see \cite{gnr15}) 
for estimating the distribution function on a compact interval in a uniform norm. More specifically, one can costruct an algorithm which allows 
estimates in the norm
\[
\left( \mathbb{E} \sup_{t \in [t_0,t_1] } \left|\P(T_s<t) -\P(\hat{T}_s<t) \right|^2 \right)^{1/2},
\] 
which guarantees the uniformity in the error for numerical estimates. \revision{This is the only Monte Carlo based approach whose 
estimates are functions, rather than finite dimensional entities, which has uniform norm as a measure of accuracy.}
\end{remark}

\section{Numerical experiments\red{, no repair}}\label{sec:examples}
\subsection{Systems and component reliability distributions}
We generated many random systems to test the MLMC reliability method proposed hereinbefore.
These random systems are generated by starting from the trivial one component system and with fixed probabilities either:
\begin{itemize}
  \item replacing a component with two components in series;
  \item replacing a component with two components in parallel;
  \item selecting two edges and inserting a `bridging' component.
\end{itemize}
This allowed us to generate a wide range of different systems and in particular an 
increasing sequence of related systems with varying numbers of component.

For all systems we consider three test cases, where the components have Weibull distributed lifetimes with shape parameter $k=0.5,1 \mbox{ or } 3$ and where the scale is chosen uniformly at random on an interval $[2,10]$. Variety in shape parameters corresponds to different applications in industry (see e.g. \cite{shooman2003reliability}). The shape parameter has a substantial effect on the corresponding density function.


\subsection{Numerical results}
We ran our algorithm 100 times for systems of different sizes with independent but differently distributed components, whose reliability is described above.  In each case we considered fixed target accuracies of $\varepsilon=2^{-4}$ and $\varepsilon=2^{-7}$, and computed the cost gains achieved for these fixed accuracies.
\subsubsection{Shape parameter $k=0.5$.}
\begin{figure}
  \centering
    \includegraphics[width=0.75\textwidth]{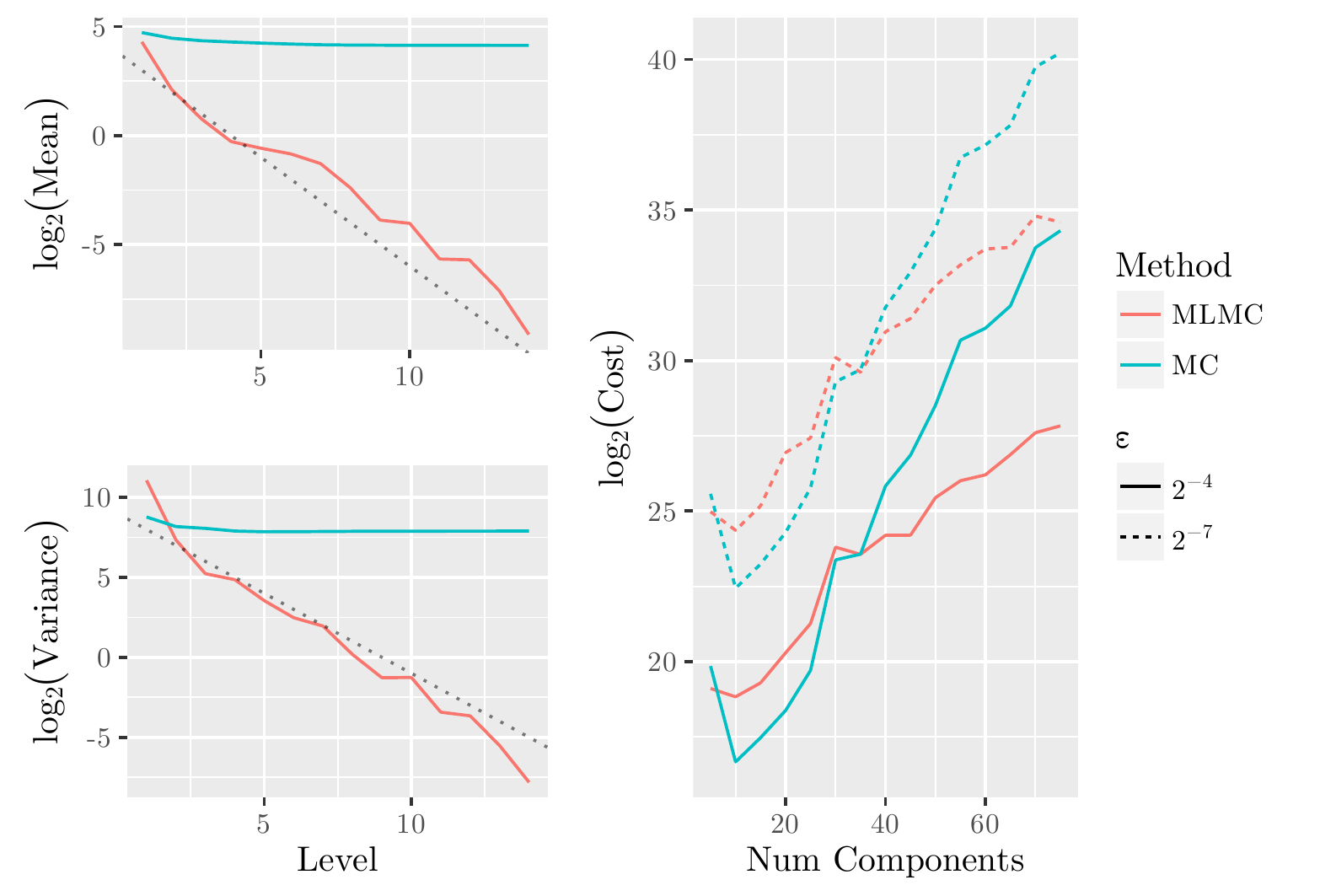}
      \caption{Left: Diagnostic tests for the largest considered system; Right: cost gains for nested randomly grown systems from 5 to 75 components, with Weibull distribution having shape parameter $k=0.5$ and uniformly distributed scale.}
      \label{fig:res_05}
\end{figure}

The left top and bottom plots on \Cref{fig:res_05} show the result of diagnostic runs, where we tested the variance and mean decay, which correspond to assumptions (3) and (1) from Theorem \ref{thm:compl} with $\beta=1$ and $\alpha=1$ respectively \red{(i.e. the slope of decay on the log-scale is $-1$)}. This indicates, that Multilevel Monte Carlo achieves the same convergence rate as traditional Monte Carlo in terms of accuracy $\varepsilon$, but can offer computational savings, due to the fact that most of the samples are calculated for very small subset of minimal cut sets. 
The right plot compares the differences in averaged costs for Multilevel Monte Carlo and standard Monte Carlo algorithms, which shows good savings even including the costs for initially selecting the cut sets for each level.
\subsubsection{Shape parameter $k=1$.}
\begin{figure}
  \centering
    \includegraphics[width=0.75\textwidth]{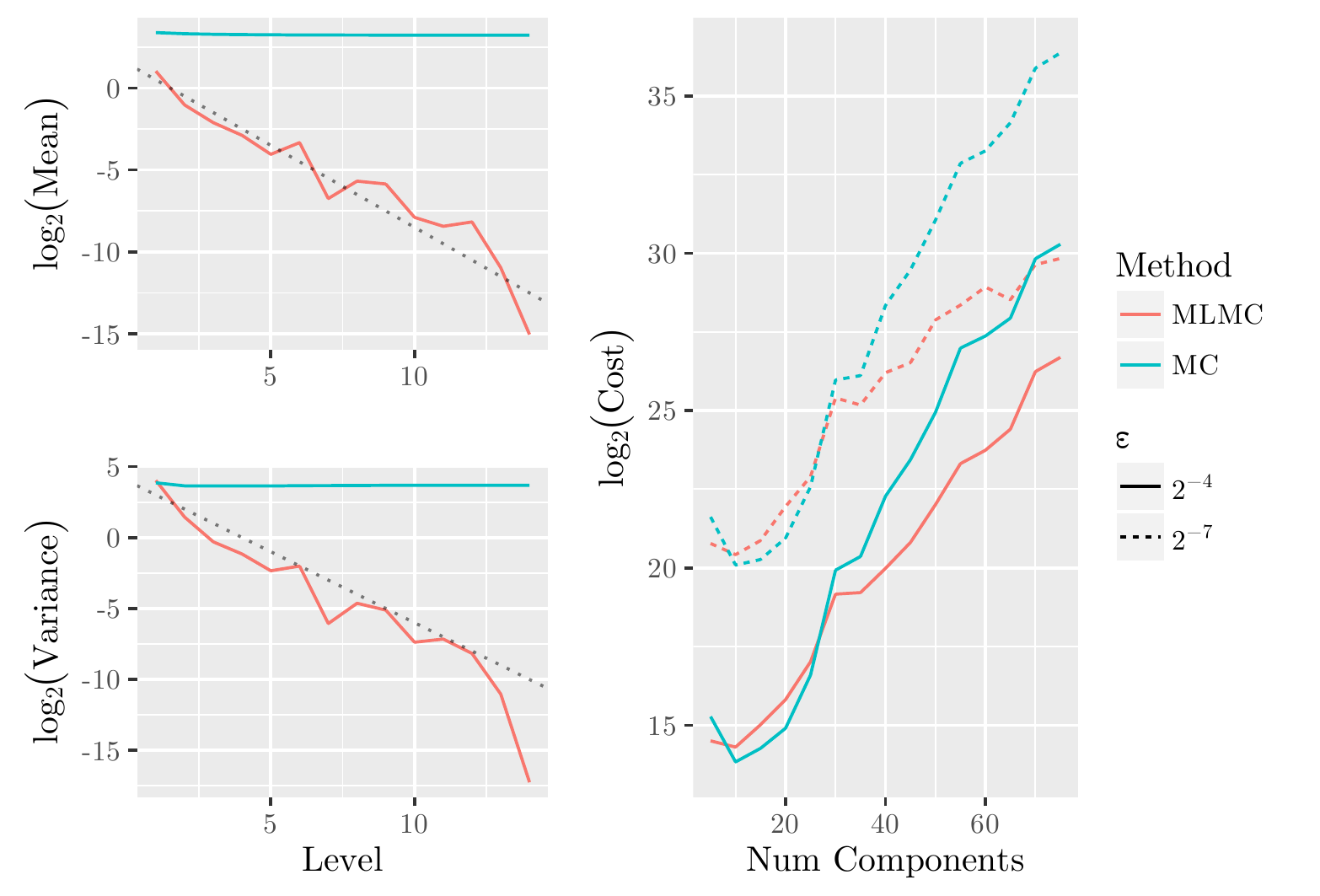}
      \caption{Left: Diagnostic tests for the largest considered system; Right: cost gains for nested randomly grown systems from 5 to 75 components, with Weibull distribution having shape parameter $k=1$ and uniformly distributed scale.}
      \label{fig:res_1}
\end{figure}

The test case with $k=1$ (\Cref{fig:res_1}) gives us almost the same mean and variance decays along with computational gains as in the case with $k=0.5$. One can see that in the last levels we see even super linear decay for the mean and variance, which indicates that the cut sets being added at those levels have very weak impact on system lifetime compared to those already chosen, which indicates good performance for the level selection algorithm.
\subsubsection{Shape parameter $k=3$.}
\begin{figure}
  \centering
    \includegraphics[width=0.75\textwidth]{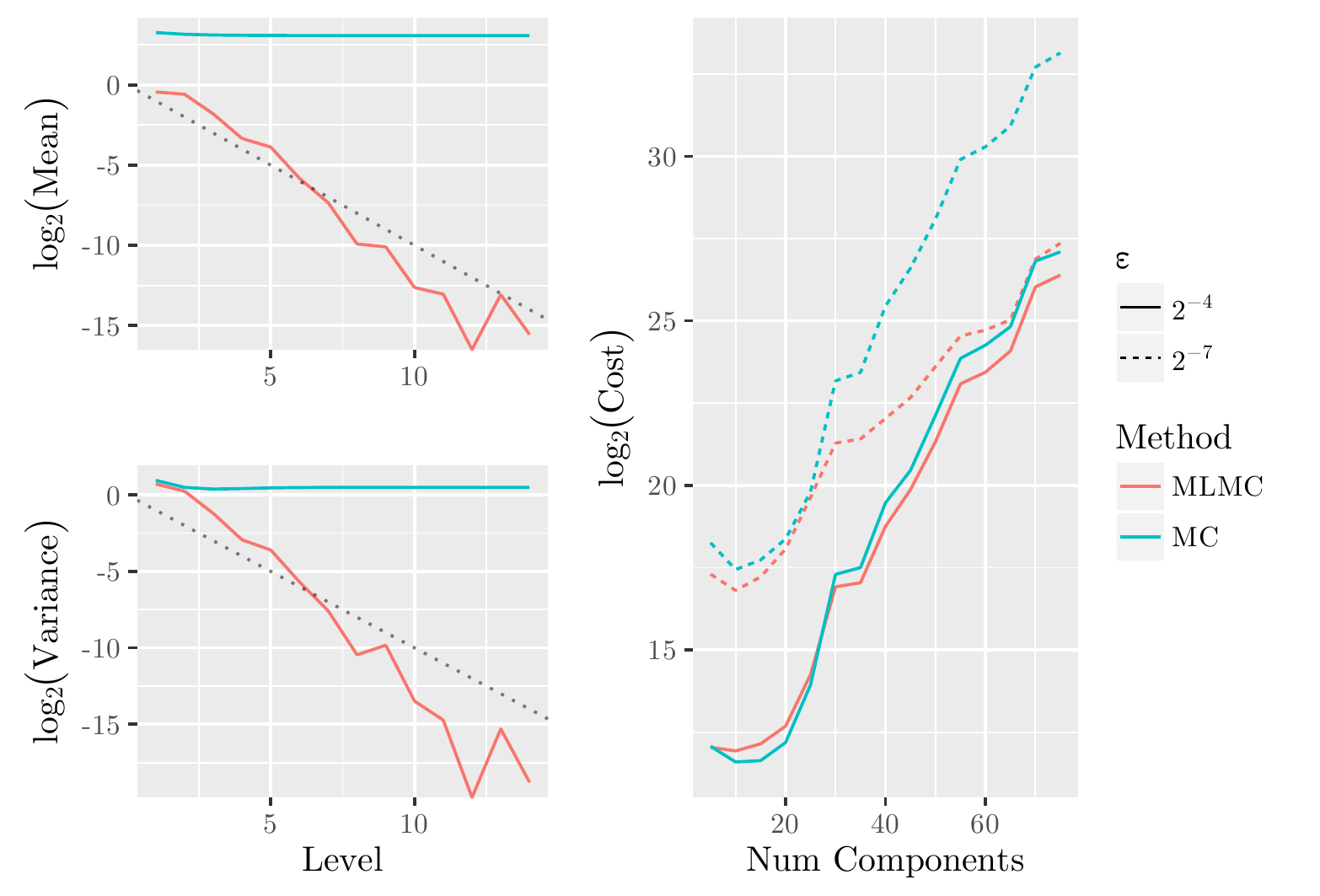}
      \caption{Left: Diagnostic tests for the largest considered system; Right: cost gains for nested randomly grown systems from 5 to 75 components, with Weibull distribution having shape parameter $k=3$ and uniformly distributed scale.}
      \label{fig:res_3}
\end{figure}

The case with $k=3$ (\Cref{fig:res_3}) shows substantial savings for $\varepsilon=2^{-7}$, as also seen in the previous examples, while still showing competitive results for $\varepsilon=2^{-4}$. The reason the gains are more modest here is that, as we had before, there is very small variance in the standard Monte Carlo and overall Multilevel Monte Carlo estimators, which puts more emphasis on the initial level selection costs which are not $\varepsilon$ dependent, but are size dependent.

\section{\red{Numerical experiments, with repair process}}\label{sec:examplesrepair}

\red{To demonstrate the generality of the method and the substantial computational benefits available in more interesting scenarios, we consider the same 70 component system as generated in \Cref{sec:examples} with a repair process included.  The components are again taken to have shape parameter $k=0.5$ and uniformly distributed scale on $[2,10]$, but now failed components are repaired according to an Exponentially distributed clock with rate $\lambda=0.05$.}

\red{Note that the final failure time of all components which lead to system failure cannot be sampled simultaneously any more, because repairs may change the state of the system en-route to ultimate failure.  Indeed, the computational complexity of sampling is substantially greater due to the need to simulate the stochastic process of failure and repair, repeatedly testing after each system state change whether the system is still functional.  Consequently, obtaining a single Monte Carlo sample may now require many passes over the collection of cut sets and moreover there is additional randomness in the runtime to simulate a single system failure time.}

\red{However, the Multilevel paradigm still applies and, as will be seen, even performs marginally better in this more complex scenario.}

\red{\subsection{Repair process}}

\begin{figure}
  \centering
    \includegraphics[scale=0.65]{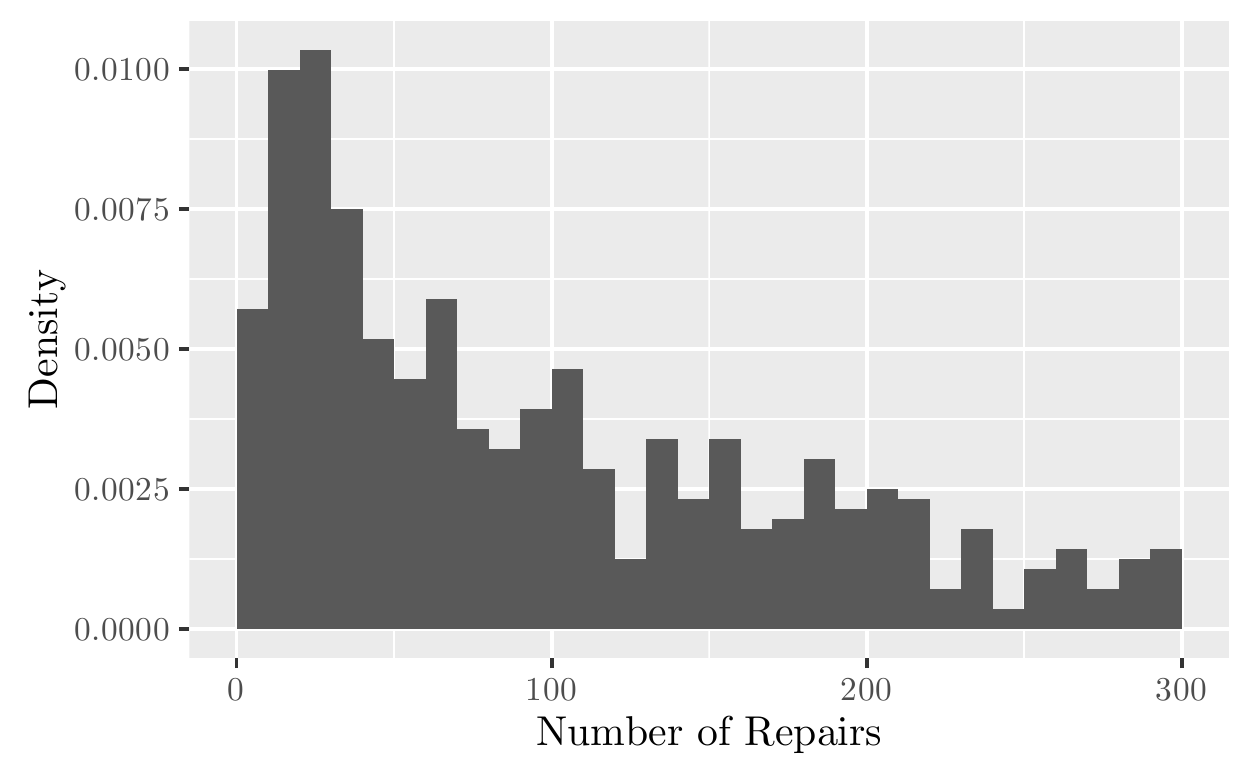}
      \caption{\red{A standard Monte Carlo estimate of the distribution of number of repairs before system failure in the example system.  Figure truncated at 300 repairs.}}
      \label{fig:repairs}
\end{figure}

\red{The repair process is taken to be Exponential($\lambda=0.05$).  Some full standard Monte Carlo runs show that this leads to a highly random number of repairs over the lifetime of the system as depicted in \Cref{fig:repairs}.  Note that this is not so much chosen to be representative of any real system, but is in fact faster repair than might be expected in order to increase the difficulty of the problem.}

\subsection{\red{Level selection for repairable systems}}
\label{sec:repair_lvlsel}

\red{The procedure described for non-repairable systems can be adapted to this case.  The only minor adjustment required is to the level selection algorithm as described in \Cref{sec:levelselection}.}

\red{Recall that the level selection procedure first involves determining a failure time for \emph{all} cut sets.  In the non-repairable case, this was simply a case of computing \cref{eq:lvl0} for a fixed collection of component simulations $t_c^{(j)}$.  However, the stochastic process of failure and repair means that $t_c^{(j)}$ depends on the cut set $C_i$ which causes failure, so that in principle for each $j$ the stochastic process should be simulated continuously until all cut sets have failed at least once, with the first failure time being recorded as $\eta_i$ for each $C_i \in \mathcal{C}$.  As such, very rare failure modes may result in essentially unbounded compute cost.}

\red{To address this, we propose simulation of the full stochastic process of failure and repair until the \emph{first} instance of failure due to a cut set.  We then simulate the conditional failure time of the still functioning components given survival to this time \emph{without} further repairs taking place.  Note that the exact behaviour beyond the initial cut set failure is therefore deemed less important: our primary goal is to establish cut sets which fail early, so exhaustive simulation is redundant.  Clearly, these simulations cannot be used in the final estimate in the way they could for the non-repairable case.}

\red{In every other way, the level selection algorithm is the exact analogue of that for the non-repairable case, with the objective being to ensure rapid decay of the mean of each level.}

\red{\subsection{Results}}

\red{\subsubsection{Diagnostics}}

\begin{figure}
  \centering
    \includegraphics[scale=0.65]{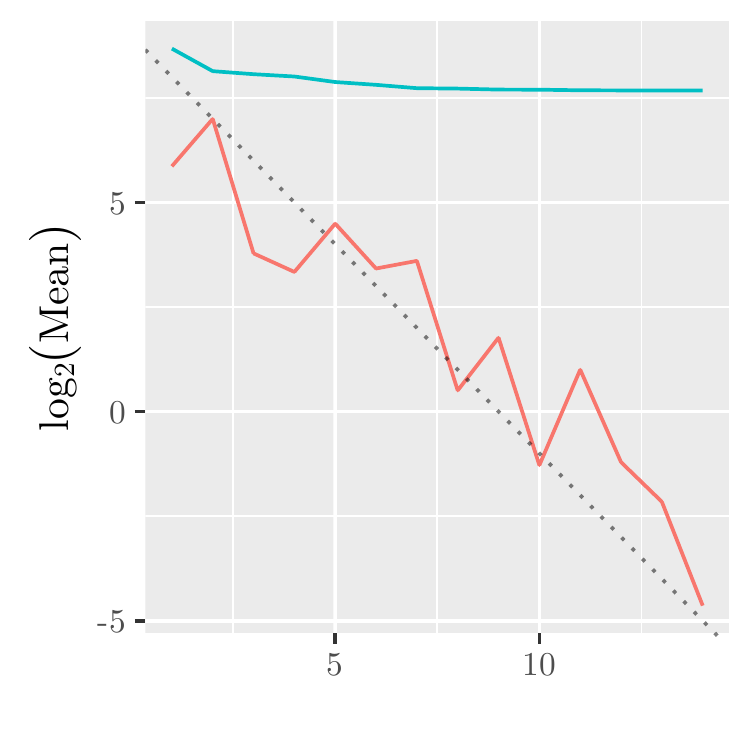} \includegraphics[scale=0.65]{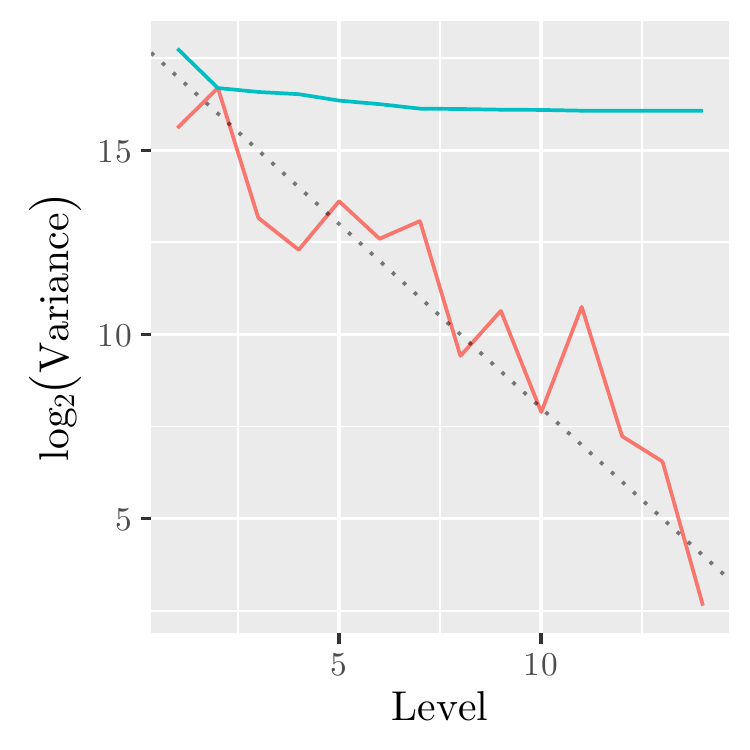}
      \caption{\red{Diagnostic plots for the repairable system example.}}
      \label{fig:repairs_diag}
\end{figure}

\red{\Cref{fig:repairs_diag} (left) shows a mean decay of order 1, which implies that the expected contribution to the system lifetime estimate halves compared to the contribution from the previous level. This strong decay, corresponding to $\alpha=1$ in \Cref{thm:compl}, means that our sorting approach does capture the influence of the cut sets, thus allowing estimation of the failure time with reasonable accuracy, without spending computational effort on evaluating `non-contributing' cut sets. Recall that the stronger the mean decay (i.e. the larger is the value $\alpha$), the fewer levels we will need, thus the higher will be the computational gain.}

\red{\Cref{fig:repairs_diag} (right) shows the variance decay and indicates the quality of the coupling. As each level is more expensive to sample than the previous one, it is desirable to sample it less often, without destroying the overall variance.  The parameter $\beta$ quantifies this in a rigorous way, in our case having $\beta \approx 1$, which implies that the variance on each level is half that on the previous level, hence halving the number of samples required.}

\red{\subsubsection{Computational cost}}

\begin{figure}
  \centering
    \includegraphics[scale=0.65]{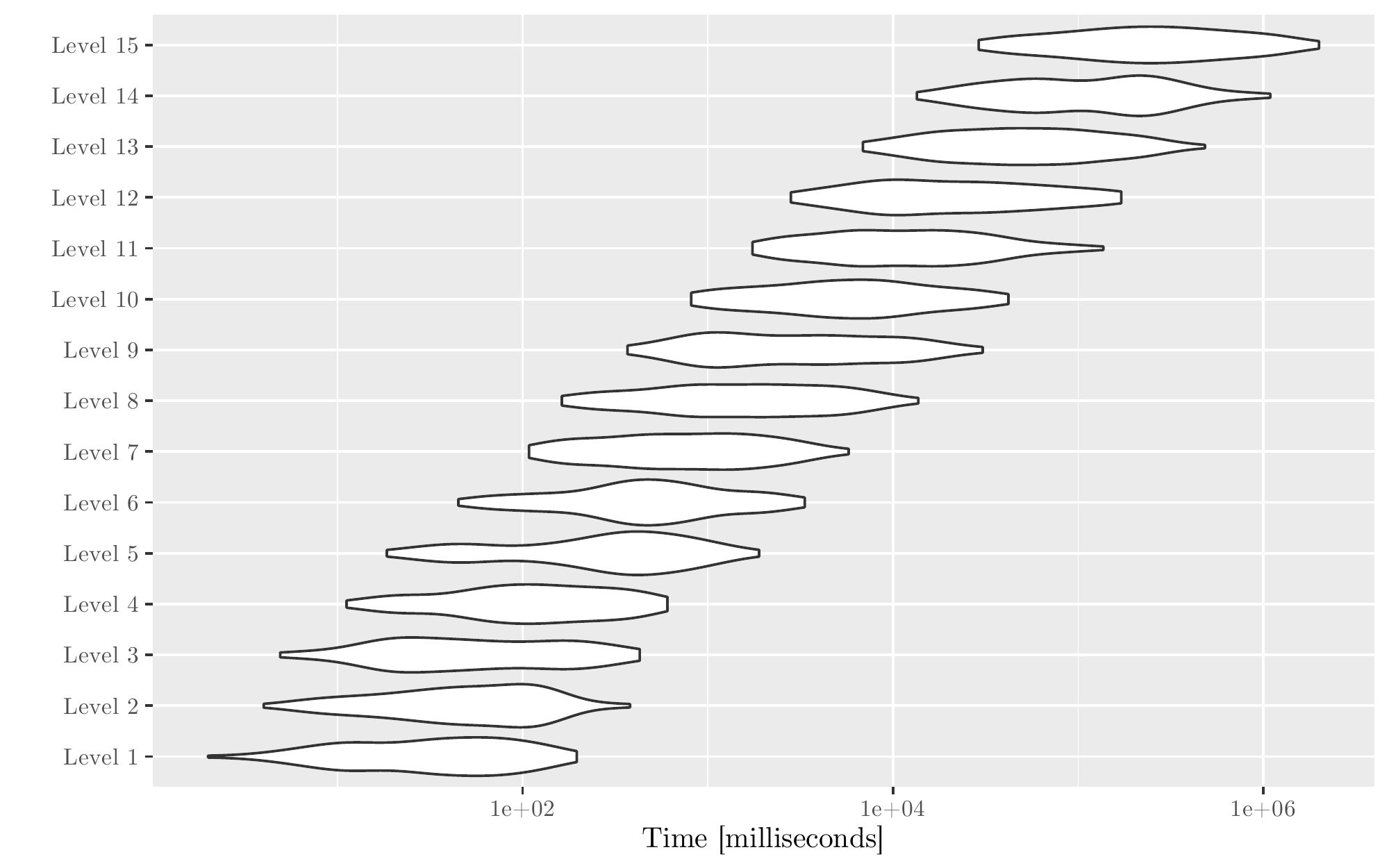}
      \caption{\red{Actual runtimes to perform a single sample on each level for the repairable example.  Note the log-scaled time axis.  Density estimates composed from 108 replicates.  Timings are for a single core of a c4.8xlarge Amazon EC2 compute instance using the AMI from \cite{AslettEC2}.}}
      \label{fig:repair_time}
\end{figure}

\red{Recall that the number of cut sets in a level was an accurate proxy for the computational cost in the non-repairable example (see p.\pageref{txt:upperbd}) and that $\gamma=1$ could then be targeted by doubling the number of cut sets in each level.}

\red{However, in the repairable case this is no longer so, because the stochastic process of failure and repair adds a random element to the simulation runtime before `system' failure, with it also depending on the cut set collection under consideration.  Therefore, \Cref{fig:repair_time} shows the distribution of empirical wall-clock runtimes to produce a single simulation for each level on a log scale, showing the desired growth in compute cost.  Regressing $\log_2(\mbox{mean runtime}) = a + \gamma \ell$ results in $\gamma\approx 0.94$.  Crucially, this means that an exponential improvement in accuracy is achieved ($\alpha=\beta=1$), but with a little below exponential increase in cost ($\gamma \approx 0.94$).  This means MLMC actually provides marginally better performance gain in the repairable case than it did in the non-repairable case (where $\alpha=\beta=\gamma=1$).}

\begin{figure}
  \centering
    \includegraphics[scale=0.65]{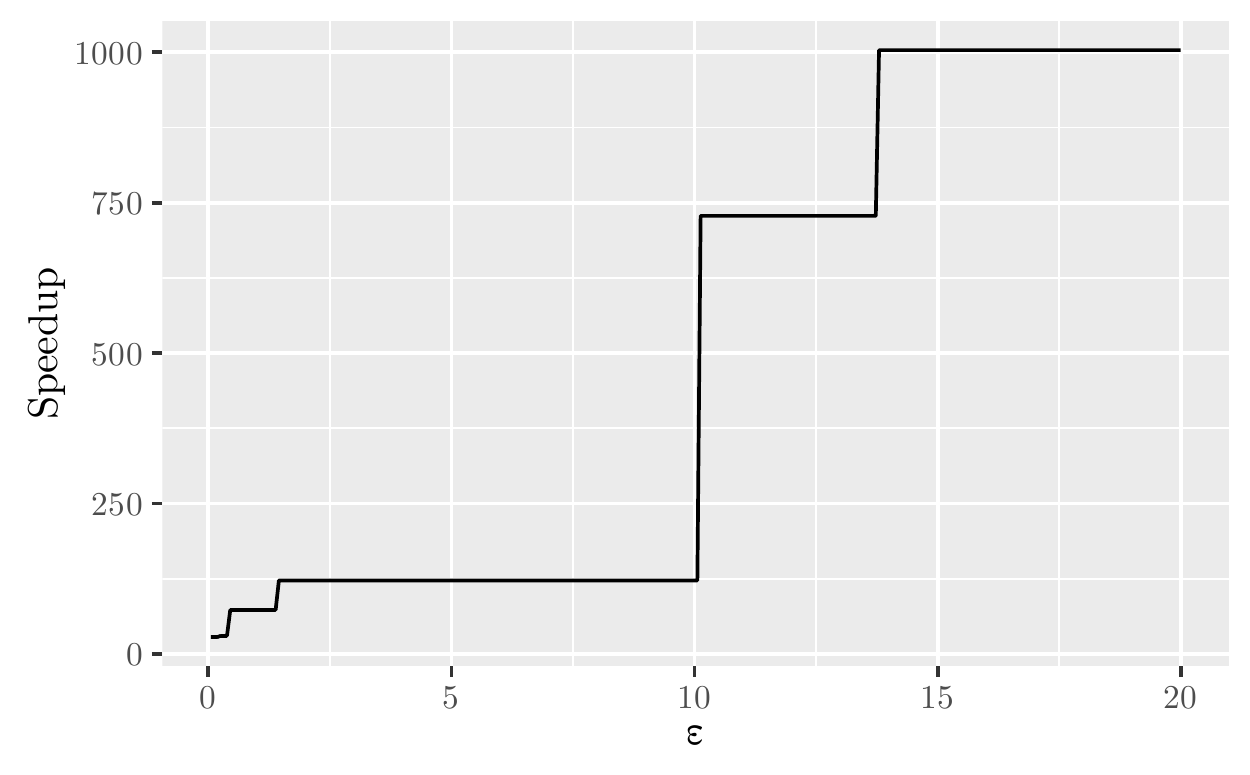}
      \caption{\red{Relative speedup of MLMC versus standard Monte Carlo to achieve estimation of expected lifetime within an accuracy of $\varepsilon$.}}
      \label{fig:repairs_speedup}
\end{figure}

\red{When computing the cost in the repairable case, we can use the empirical mean compute time for level $\ell$, $\kappa_\ell$ say, instead of the number of cut sets in \cref{eq:cost:mlmc}.  Then, for a target accuracy $\varepsilon$, the speedup provided by MLMC is characterised by:}
\begin{equation*}
  \mbox{Speedup} = \frac{\varepsilon^{-2} \V(\tau_i) \kappa_L}{\varepsilon^{-2} \left( \sum_{\ell=0}^{L_\varepsilon} \sqrt{\V(Y_\ell) \kappa_\ell} \right)^2}
\end{equation*}
\red{where $L_\varepsilon$ is the earliest level with mean less than $\varepsilon$.  For varying $\varepsilon$ this is depicted in \Cref{fig:repairs_speedup}.  For coarse estimates, speedups of upto $1,000$ times can be achieved.  Note that the expected system lifetime is $\approx 205$ with high variance $\approx 69,000$ --- this was chosen as an extreme example to test MLMCs ability to handle very difficult estimation problems.}

\red{To put this in perspective, the time to achieve a Monte Carlo estimate to accuracy $\varepsilon=1.5$ would take about 141 days on a single CPU core (or nearly 4 days using all cores of the c4.8xlarge Amazon EC2 instance used for testing).  Multilevel Monte Carlo would achieve the same accuracy in 1 day, 4 hours on a single CPU core (or just 45 minutes using all cores of a c4.8xlarge).}

%

\section{Conclusion and future work}\label{sec:conc}
\textcolor{red}{We have presented an exciting new application for the Multilevel paradigm for estimating the reliability of systems, which speeds up traditional Monte Carlo estimation of system lifetimes and provides a approach which can easily generalise to other reliability problems which involve cut (path) sets.  We have demonstrated that the proposed approach to using MLMC in reliability problems achieves the strong mean and variance decay required to enable truncation of the number of levels which must be simulated.  This is a very desirable feature for MLMC, as the standard approach has to go through all the cut sets for each sample, regardless of target accuracy.  Moreover, we have demonstrated that harder problems (such as repairable systems) achieve slightly greater gains through just below exponential growth in the cost of simulating levels ($\gamma\approx 0.94$) while still having exponential mean and variance decay ($\alpha=\beta=1$)}

Unlike classic MLMC implementations, where one considers different approximations of a certain stochastic process wherein all of them are biased, here we introduce approximations based on sorting the minimal cut sets in a special way, which are biased, but less costly to simulate.
The numerical experiments show substantial savings for large systems and are promising for further study of reliability and structural optimisation. \red{Indeed, the ease of extending from non-repairable to a repairable setting shows the flexibility of the approach and we anticipate it would be likewise straight-forward to incorporate shock and stress processes in a similar manner, benefitting ever more greatly from the acceleration offered by MLMC.} \textcolor{red}{Our own future work will include the extension of our approach in the spirit of \cite{gnr15}, and expanding the applicability of the Multilevel paradigm to other algorithms established in the reliability community.}

\section*{Acknowledgments}

\red{The authors would like to thank four reviewers for comments which greatly improved the presentation of this paper.}

Louis Aslett is supported by the i-like programme grant (EPSRC grant reference number EP/K014463/1 http://www.i-like.org.uk).  Tigran Nagapetyan and Sebastian Vollmer received funding from EPSRC Grant EP/N000188/1.

\section*{References}
\bibliographystyle{plain}
\bibliography{mlmc} 

\end{document}